\documentclass[showpacs,preprintnumbers,amsmath,amssymb]{revtex4}

\usepackage{epsfig}
\usepackage{graphicx}
\usepackage{dcolumn}
\usepackage{bm}

\begin{document}

\title{Grouping time series by pairwise measures of redundancy}
\author{D. Marinazzo$^{1}$, W. Liao$^{2}$, M. Pellicoro$^{3,4,5}$, and S.
Stramaglia$^{3,4,5}$}

 \affiliation{$^1$ Laboratoire de Neurophysique et Neurophysiologie, Universit\'e Paris Descartes, Paris, France\\}
 \affiliation{$^2$
Key Laboratory for NeuroInformation of Ministry of Education, School
of Life Sciences and Technology, University of Electronic Science
and Technology of China, China
\\}
 \affiliation{$^3$ Istituto Nazionale di Fisica
Nucleare, Sezione di Bari, Italy\\} \affiliation{$^4$ Dipartimento
di Fisica, University of Bari, Italy\\} \affiliation{$^5$
TIRES-Center of Innovative Technologies for Signal Detection and
Processing, Universit\`a di Bari,  Italy\\}

\date{\today}

\begin{abstract}
A novel approach is proposed to group redundant time series in the
frame of causality. It assumes that (i) the dynamics of the system
can be described using just a small number of characteristic modes,
and that (ii) a pairwise measure of redundancy is sufficient to
elicit the presence of correlated degrees of freedom. We show the
application of the proposed approach on fMRI data from a resting
human brain and gene expression profiles from HeLa cell culture.
\end{abstract}

\maketitle Over the last years the interaction structure of many
complex systems has been mapped in terms of graphs, which can be
characterized using tools of statistical physics
 \cite{barabasi}. Dynamical networks model physical and
biological behavior in many applications; examples range from
networks of neurons \cite{netnn}, Josephson junctions arrays
\cite{jose} to genetic networks \cite{gennet}, protein interaction
nets \cite{protein} and metabolic networks \cite{metabolic}.
Synchronization in dynamical networks is influenced by the topology
of the network \cite{bocca}. The inference of dynamical networks is
related to the estimation, from data, of the flow of information
between variables. Two major approaches are commonly used to
estimate the information flow between variables, transfer entropy
\cite{schreiber} and Granger causality \cite{granger}.

An important notion in information theory is the redundancy in a
group of variables, formalized in \cite{palus} as a generalization
of the mutual information. A formalism to recognize redundant and
synergetic variables in neuronal ensembles has been proposed in
\cite{sch} and generalized in \cite{bettencourt}. Recently a
quantitative definition to recognize redundancy and synergy in the
frame of causality has been provided \cite{noipre10} and it has been
shown that the maximization of the total causality, over all the
possible partitions of variables, is connected to the detection of
groups of redundant variables; the search over all the partitions is
unfeasible but for small systems. We remark that the information
theoretic treatments of groups of correlated degrees of freedom can
reveal their functional roles in complex systems. The purpose of
this work is to propose a simple approach to find groups of causally
redundant variables (groups of variables sharing the same
information about the future of the system), which can be applied
also to large systems. The main assumption underlying our approach
is that the essential features of the dynamics of the system under
consideration are captured using just a small number of
characteristic modes. Hence we use principal components analysis to
obtain a compressed representation of the future state of the
system. Then, we introduce a pairwise measure of the redundancy
w.r.t. the prediction of the next configuration of the modes, thus
obtaining a weighted graph. Finally, by maximizing the modularity
\cite{bocca}, we find the natural modules of this weighted graph and
identify them with the groups of redundant variables. In the
following section we describe the method. In section II we describe
the application of the method to fMRI data, and in section III to a
gene expression data-set. Some conclusions are drawn in section IV.

\section{Method}
Let us consider $n$ time series $\{x_i(t)\}_{i=1,\ldots,n}$; after a
linear transformation, we may assume all the time series to be
normalized and with zero mean. The lagged times series are denoted
$X_i(t)= x_i (t-1)$. We make the hypothesis that the dynamics of the
system under consideration may be described in terms of a few modes,
and that these modes may be extracted by principal components
analysis, as follows. Calling ${\bf x}$ the $n\times T$ matrix with
elements $x_i(t)$, we denote
$\{u_\alpha(t)\}_{\alpha=1,\ldots,n_\lambda}$ the (normalized)
eigenvectors of the matrix ${\bf x}^\top {\bf x}$ corresponding to
the largest $n_\lambda$ eigenvalues. The T-dimensional vectors
$u_\alpha(t)$ summarize the dynamics of the system; the lagged
correlations of the system determine to what extent the modes $u$
may be predicted on the basis of the $X_i(t)$ variables.

Preliminarily, we select the variables which are significatively
correlated with the modes $u$. For each $i$ and each $\alpha$ we
evaluate the probability $p_{i\alpha}$ that the correlation between
$X_i$ and $u_\alpha$ is due to chance, obtained by Student's t test.
We compare $p_{i\alpha}$ with the $5\%$ confidence level after
Bonferroni correction (the threshold is $0.05/(n\times n_\lambda)$)
and retain only those variables $X_i$ which are significatively
correlated with at least one mode. The variables thus selected will
be denoted  $\{Y_i(t)\}_{i=1,\ldots,N}$, $N$ being their
cardinality.

The second step of the present approach is the introduction of a
bivariate measure of redundancy, as follows. For each pair of
variables $Y_i$ and $Y_j$, we denote $P_i$ the projector onto the
one-dimensional space spanned by $Y_i$ and $P_j$ the projector onto
the space corresponding to $Y_j$; $P_{ij}$ is the projector onto the
bi-dimensional space spanned by $Y_i$ and $Y_j$. Then, we define:
\begin{equation}\label{redundancy}
c_{ij}=\sum_{\alpha=1}^{n_\lambda}\left( ||P_i u_\alpha ||^2 + ||P_j
u_\alpha ||^2 -||P_{ij} u_\alpha ||^2\right);
\end{equation}
according to the discussion in \cite{noipre10}, $c_{ij}$ is positive
(negative) if variables $Y_i$ and $Y_j$ are redundant (synergetic)
w.r.t. the prediction of the future of the system. In other words,
if $Y_i$ and $Y_j$ share the same information about $u$, then
$c_{ij}$ is positive.

In the third step, the matrix $c_{ij}$ is used to construct a
weighted graph of $N$ nodes, the weight of each link measuring the
degree of redundancy between the two variables connected by that
link. By maximization of the modularity \cite{newman}, the number of
modules, as well as their content, is extracted from the weighted
graph. Each module is recognized as a group of variables sharing the
same information about the future of the system.

\begin{figure}[ht!]
\begin{center}
\epsfig{file=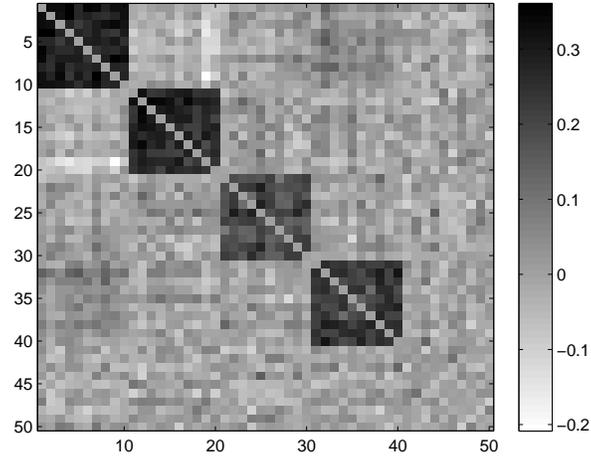,height=7.cm}
\end{center}
\caption{{\small The correlation matrix of the simulated example,
showing four groups of variables correlated at equal times.
\label{fig1}}}
\end{figure}

As an example we report the following example. Let us consider the
following autoregressive system:
\begin{eqnarray}
\begin{array}{ll}
\psi_{t}&=0.6 \eta_{t-1}+0.1 \xi^1_{t}\\
\eta_{t}&=0.6 \psi_{t-1}+0.1 \xi^2_{t},
\end{array}
\label{map}
\end{eqnarray}
where $\xi$ are i.i.d. unit variance Gaussian variables. By
construction, $\psi$ is caused by $\eta$ and viceversa. A system of
$50$ time series is constructed as follows. For $i=1,\ldots,10$:
\begin{eqnarray}
\begin{array}{cl}
x_i (t)&=\psi_{t} + 0.2 \rho^i_{t},\\
x_{10+i} (t)&=\eta_{t} + 0.2 \rho^{10+i}_{t},\\
x_{20+i} (t)&=\xi^3_{t} + 0.2 \rho^{20+i}_{t},\\
x_{30+i} (t)&=\xi^4_{t} + 0.2 \rho^{30+i}_{t},\\
x_{40+i} (t)&=\rho^{40+i}_{t},\\
\end{array}
\label{map1}
\end{eqnarray}
where $\rho$ and $\xi$ are i.i.d. unit variance Gaussian variables.
Starting from a random initial configuration, the above equations
are iterated and, after discarding the initial transient regime,
$n_s$ consecutive samples of the system are stored for further
analysis. Note that the first ten variables share the same
information corresponding to $\psi$, whilst the second ten variables
share the information of $\eta$. The variables $x_i$, with
$i=21,\ldots,30$, form a group of variables with correlations at
equal times, similarly to the group of variables with
$i=31,\ldots,40$. The variables $x_i$, with $i=41,\ldots,50$,
correspond to pure noise. In figure (\ref{fig1}) the equal-times
correlations of the system are depicted, for a typical case with
$n_s =500$, showing four groups of correlated variables. We perform
the principal components analysis and retain a variable number
$n_\lambda$ of modes for the analysis.

\begin{figure}[ht!]
\begin{center}
\epsfig{file=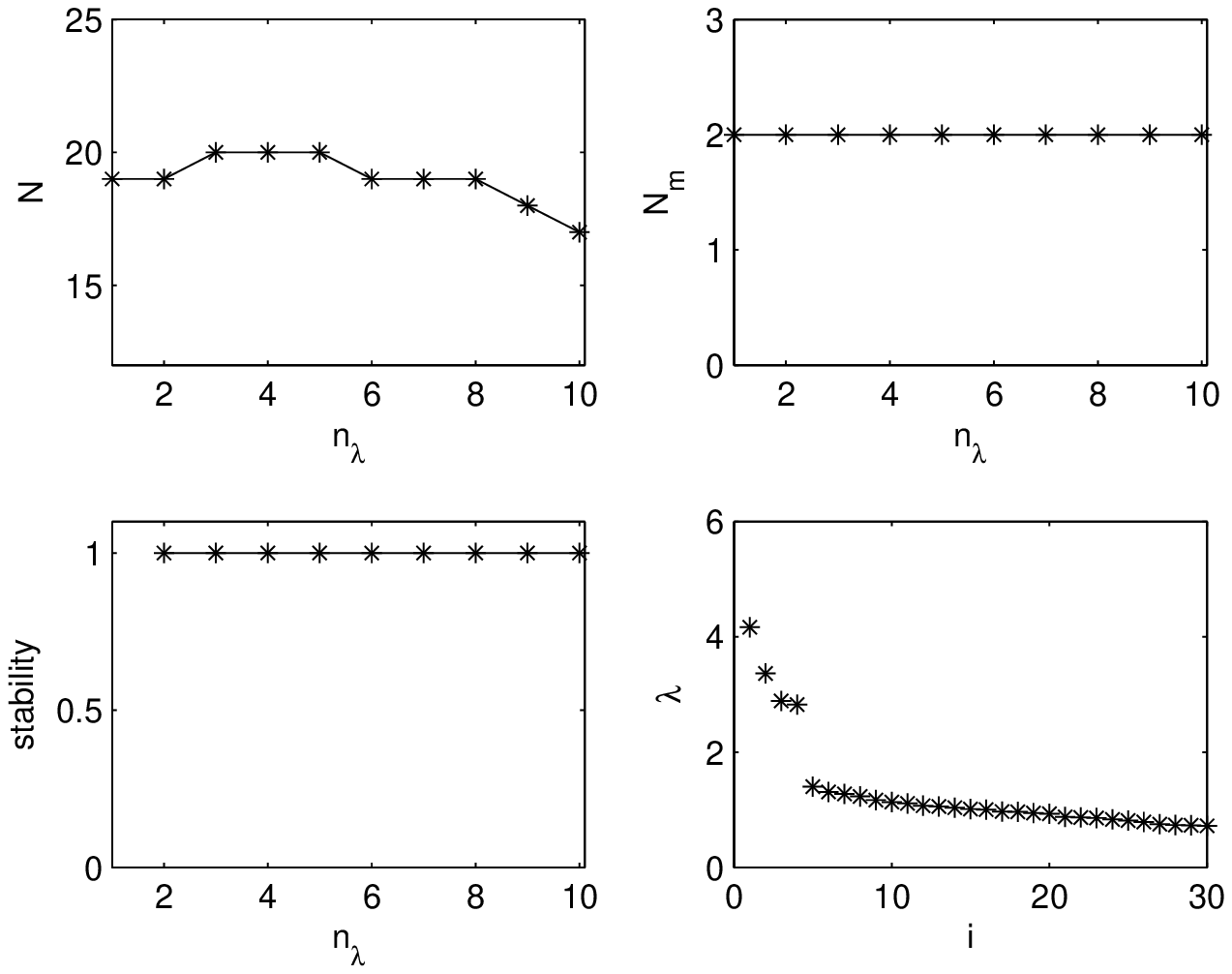,height=7.cm}
\end{center}
\caption{{\small (Top-left) Concerning the simulated example, the
number of selected variables $N$ is plotted versus $n_\lambda$, the
number of modes. (Top-right) The number of modules, obtained by
modularity maximization, of the matrix $c_{ij}$, whose elements
measure the pairwise redundancy. (Bottom-left) The measure of the
stability of the partition, going from $n_\lambda -1$ to
$n_\lambda$, is plotted versus $n_\lambda$. (Bottom-right) The
eigenvalues of the matrix $x^\top x$ are depicted. \label{fig2}}}
\end{figure}

In figure (\ref{fig2}), top-left, we depict $N$, the number of
selected variables, as a function of $n_\lambda$. For
$n_\lambda=3,4,5$, twenty variables ($x_i$ with $i=1,\ldots,20$) are
selected; nineteen variables for $n_\lambda=1,2,6,7,8$. Then, for
each value of $n_\lambda$, the quantities $c_{ij}$ are evaluated. We
find that, in this example, the matrix $c_{ij}$ is non-negative and
can be treated as a weighted graph.

In figure (\ref{fig2}), top-right, we plot the number of modules
$N_m$ we find by applying the method described in \cite{newman} to
the matrix $c_{ij}$; the method correctly recognizes the two modules
for each value of $n_\lambda$.

In figure (\ref{fig2}), bottom-left, we plot a measure of the
stability of the partition while going from $n_\lambda -1$ to
$n_\lambda$, defined as follows. We consider all the pairs of
variables that are selected both in correspondence of $n_\lambda -1$
and $n_\lambda$. The stability is one minus the fraction of pairs
such that the variables are recognized to be in the same module in
one instance and in different modules in the other instance. In this
case the stability is always one; when the method is applied to real
data, the stability curve may be helpful to fix the optimal number
of modes $n_\lambda$.

Finally, in figure (\ref{fig2}), bottom-right, the eigenvalues of
the matrix $x^\top x$ are depicted. In this case it is clear that
the optimal number of modes is four.

\begin{figure}[ht!]
\begin{center}
\epsfig{file=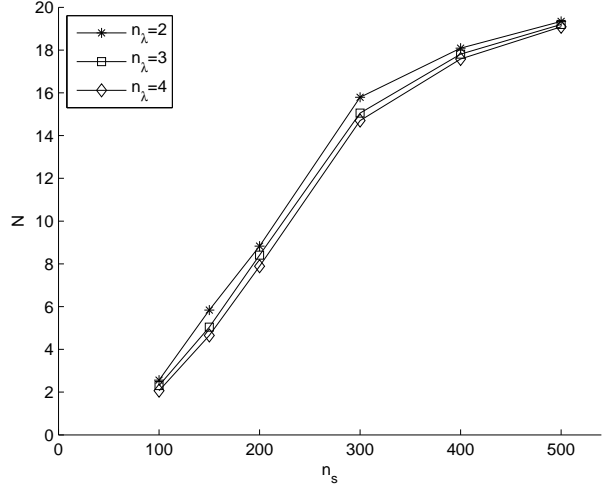,height=7.cm}
\end{center}
\caption{{\small The number of selected variables, for the simulated
example, is depicted as a function of the number of samples $n_s$
for $n_\lambda= 2,3,4$. \label{fig3}}}
\end{figure}

We remark that a suitable number of samples is needed to obtain
reliable results. In figure (\ref{fig3}) we depict the number of
selected variables, for this example, as a function of $n_s$ for
three choices of $n_\lambda$: it vanishes as the number of samples
decreases.

\section{Modular organization of brain activity}
The fMRI signal  can be regarded as a proxy for the underlying
neural activity. Remote regions of the brain do not operate in
isolation and there is a growing interest in studying the
interactions and connectivity patterns between these regions, which
have been investigated by independent component analysis
\cite{liao}, principal components analysis \cite{lohmann}  and other
approaches. Temporal and spatial functional networks, corresponding
to spontaneous brain activity in humans, were derived in \cite{he}
on the basis of the equal-time correlation matrix. Modularity in the
resting state of the human brain has also been studied in
\cite{ferrarini,meunier,van}. The connectivity structure of brain
networks extracted from spontaneous activity signals of healthy
subjects and epileptic patients has been analyzed in
\cite{latora,liao1}.

Here we consider fMRI data from a subject in resting conditions,
with sampling frequency 1 Hz, and number of samples equal to 500. A
prior brain atlas is utilized to parcellate the brain into ninety
cortical and subcortical regions, and a single time series is
associated to each region. All the ninety time series are then
band-passed in the range 0.01-0.08 Hz.

\begin{figure}[ht!]
\begin{center}
\epsfig{file=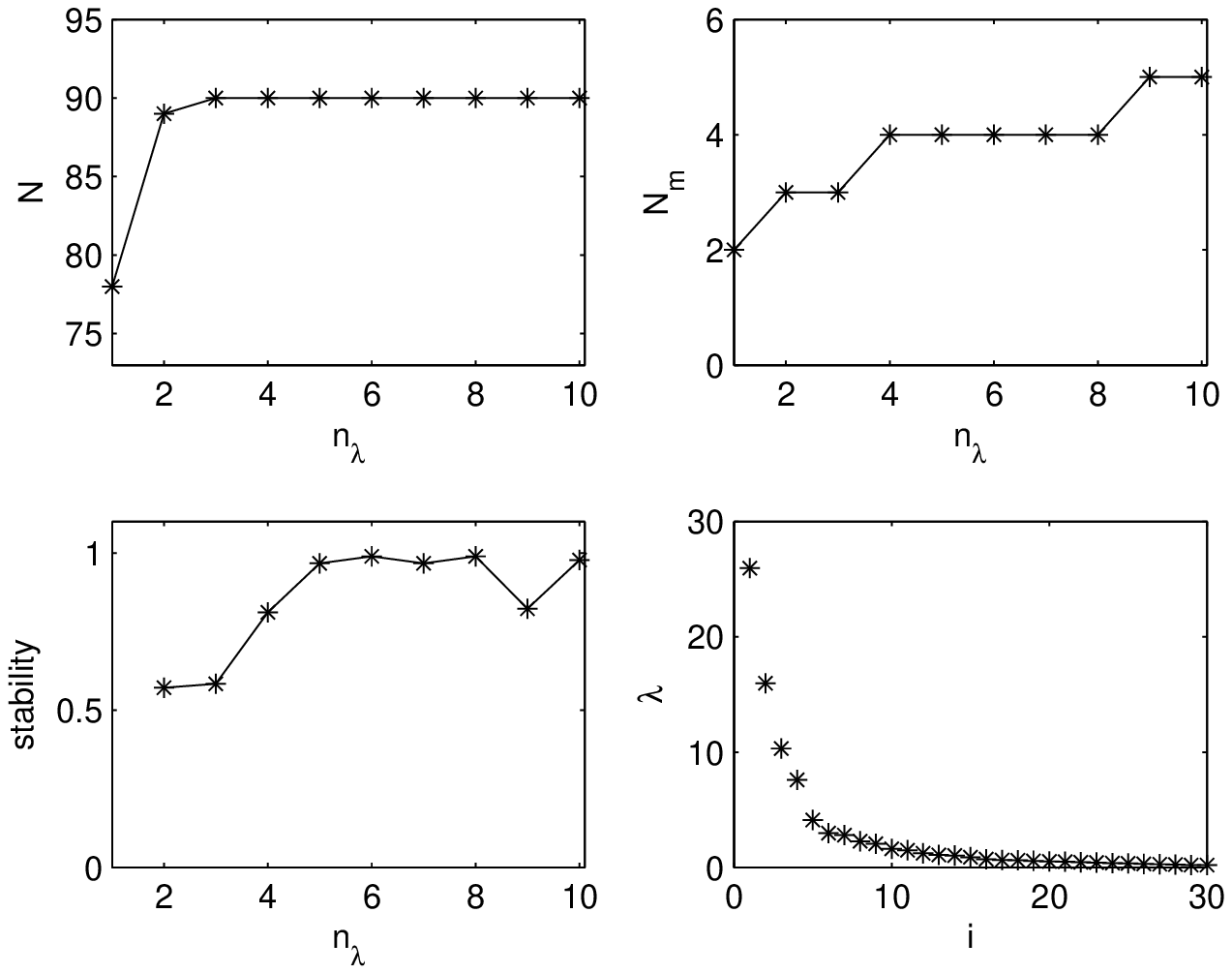,height=7.cm}
\end{center}
\caption{{\small (Top-left) Concerning the fMRI application, the
number of selected regions $N$ is plotted versus $n_\lambda$, the
number of modes. (Top-right) The number of modules, obtained by
modularity maximization, of the matrix $c_{ij}$, whose elements
measure the pairwise redundancy. (Bottom-left) The measure of the
stability of the partition, going from $n_\lambda -1$ to
$n_\lambda$, is plotted versus $n_\lambda$. (Bottom-right) The
eigenvalues of the matrix $x^\top x$ are depicted. \label{fig4}}}
\end{figure}

In figure (\ref{fig4}), top-left, we depict $N$, the number of
selected variables, as a function of $n_\lambda$. For $n_\lambda
>3$, all the ninety regions are recognized as influencing the future
of the system. For each value of $n_\lambda$, the quantities
$c_{ij}$ are evaluated. In figure (\ref{fig4}), top-right, the
number of modules $N_m$ we find by applying the method described in
\cite{newman} to the matrix $c_{ij}$ is depicted; this plot suggests
the presence of four modules for $4< n_\lambda < 8$. These values
are the most stable, as it is clear from figure (\ref{fig2}),
bottom-left, where we plot the measure of the stability of the
partition while going from $n_\lambda -1$ to $n_\lambda$. It follows
that the optimal value of $n_\lambda$ is four, corresponding to a
graph structure with four modules and modularity equal to 0.3. We
find that module 1 includes brain regions from ventral medial
frontal cortices which are primarily specialized for anterior
default mode network, module 2 is typical referred to as posterior
default mode network, module 3 mainly corresponds to executive
control network and module 4 refers to the subcortical network. In
figure (\ref{fig4}), bottom-right, the eigenvalues of the matrix
$x^\top x$ are depicted.

\begin{figure}[ht!]
\begin{center}
\epsfig{file=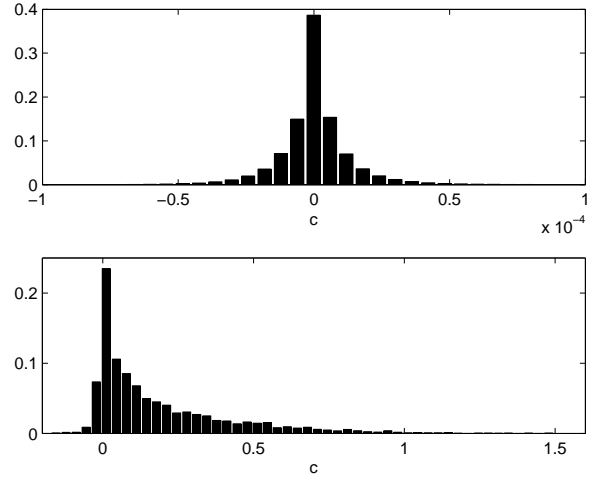,height=7.cm}
\end{center}
\caption{{\small The histogram of the values of the pairwise
redundancy $c_{ij}$, in fMRI example (bottom), and choosing randomly
the modes $u$ (top) \label{fig5}}}
\end{figure}

It is worth comparing the histogram of the values of $c_{ij}$, in
this example (figure (\ref{fig5})-bottom), with those corresponding
to a random choice of the modes $u$ (figure (\ref{fig5})-top). In
the random case, the pairs of variables are either redundant or
synergetic, and the typical values of $c$ are very small. On the
data set at hand, the magnitude of the values of $c$'s is much
higher and variables are mostly redundant; indeed the $c$'s are
negative (with small absolute value) only for a few pairs of
regions. We remark that the presence of a few small and negative
weights does not influence significantly the output from the
modularity algorithm of \cite{newman}: the output does not change if
all $c$'s with absolute value less than a threshold are set to zero
(the threshold being chosen so that all the elements of the
resulting matrix are nonnegative).

\begin{figure}[ht!]
\begin{center}
\epsfig{file=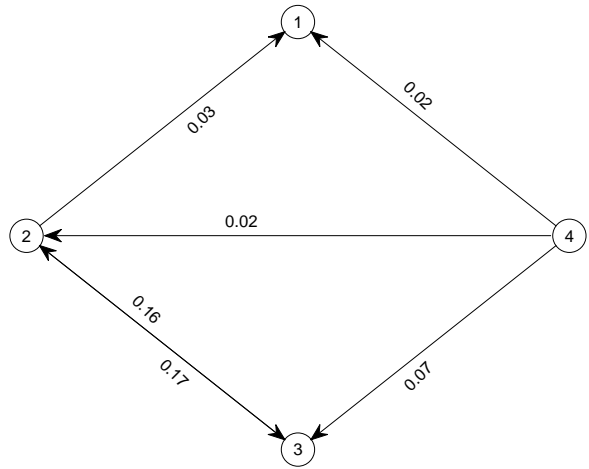,height=7.cm}
\end{center}
\caption{{\small The causalities between the four modules of the
fMRI application. \label{fig6}}}
\end{figure}

Averaging the time series belonging to each module, we obtain four
time series and we evaluate the causalities between them: the result
is displayed in figure (\ref{fig6}). It is interesting to observe
that module 4 influences all the three other modules but is not
influenced by them, it is an out-degree hub; this is consistent with
the fact that it corresponds to subcortical brain. Another striking
feature is the clear interdependencies between modules 2 and 3. The
reliability of this pattern needs to be assessed on a large
population of subjects.
\section{HeLa gene expression regulatory network} HeLa \cite{hela}
is a famous cell culture, isolated from a human uterine cervical
carcinoma in 1951. HeLa cells have somehow acquired cellular
immortality, in that the normal mechanisms of programmed cell death
after a certain number of divisions have somehow been switched off.
 We consider the HeLa cell gene expression data
of \cite{fujita}. Data corresponds to $94$ genes and $48$ time
points, with an hour interval separating two successive readings
(the HeLa cell cycle lasts 16 hours). The 94 genes were selected,
from the full data set described in \cite{whitfield}, on the basis
of the association with cell cycle regulation and tumor development.
This data has been also considered in \cite{noipre}. The static
correlation analysis between time series, which is the result of
regulation mechanisms with time scales faster than the sampling
rate, revealed a highly related network with the presence of two
modules: the first module was recognized as corresponding to the
regulatory network of the transcriptional factor NFkB \cite{nfkb},
whilst the second module appeared to be orchestrated by
transcriptional factors p53 and STAT3. Use of bivariate Granger
causality, in \cite{noipre}, has put in evidence 19 causality
relationships acting on the time scale of one hour, all involving
genes playing some role in processes related to tumor development.

As stated in \cite{maritan}, fundamental patterns underlie gene
expression profiles.  This suggest the use of the proposed approach
on gene expression time series. In figure (\ref{fig6}) we describe
the application of the proposed approach on the HeLa data-set. The
stable partition corresponds to $n_\lambda =4$ and consists of two
modules of 9 and 7 genes (the modularity is 0.1). The first module
is characterized by the transcriptional factor NFkB and consists of
NFkB, MCP-1, ICAM-1, Bcl-XL, IAP, A20, c-myc, TSP1, and Mcl-1. The
second module is related to the transcriptional factor JunB, known
to be a regulator of life and death of cells \cite{junb}, and
consists of JunB, IL-6, IkappaBa, P21, Noxa, c-jun and NRBP.
Averaging the time series belonging to each module, and evaluating
the causality between the two time series thus obtained, we obtain a
relevant (0.145) causality of the first module on the second one. We
note that all the sixteen genes selected by our approach were
recognized as interacting in \cite{fujita}; nine of them were
involved also in the causalities described in \cite{noipre}. It is
not surprising that different methods, on the same data set, provide
slightly different results: currently available data size and data
quality make the reconstruction of gene regulatory networks from
gene expression data a challenge. In figure (\ref{fig5})-bottom) we
depict the histogram of the values of $c$ on this data-set, with
those corresponding to choosing randomly the modes $u$ (figure
(\ref{fig5})-top). On this data, the values of $c$ that we obtain
are significatively greater than those one finds in the random case,
and all the pairs are redundant.

\section{Conclusions}
Grouping redundant time series reveals their functional roles in
complex systems. In the frame of causal approaches, grouping
redundant time series may reflect directed influence of one group
over another. In this work we have proposed a novel approach which
assumes that (i) the dynamics of the system can be described using
just a small number of characteristic modes, and that (ii) a
pairwise measure of redundancy is sufficient to elicit the presence
of correlated degrees of freedom. Grouping is provided by the
identification of the modules of the weighted graph of redundancies.
The method may be seen as an alternative to the analysis of
\cite{noipre10}, which can be applied also for large systems. We
have shown the effectiveness of the proposed approach in two
applications, the analysis of fMRI data and the analysis of gene
expression data. In both these applications usually linear
interactions are sought for, and only lags of 1 are considered,
therefore here we limited to consider a linear pairwise measure of
redundancy, and considered only lags of 1. The generalization of the
pairwise redundancy measure, here introduced, to the nonlinear case
and to lags of higher order is matter for further work, along the
lines described in \cite{noipre10}.


\begin{figure}[ht!]
\begin{center}
\epsfig{file=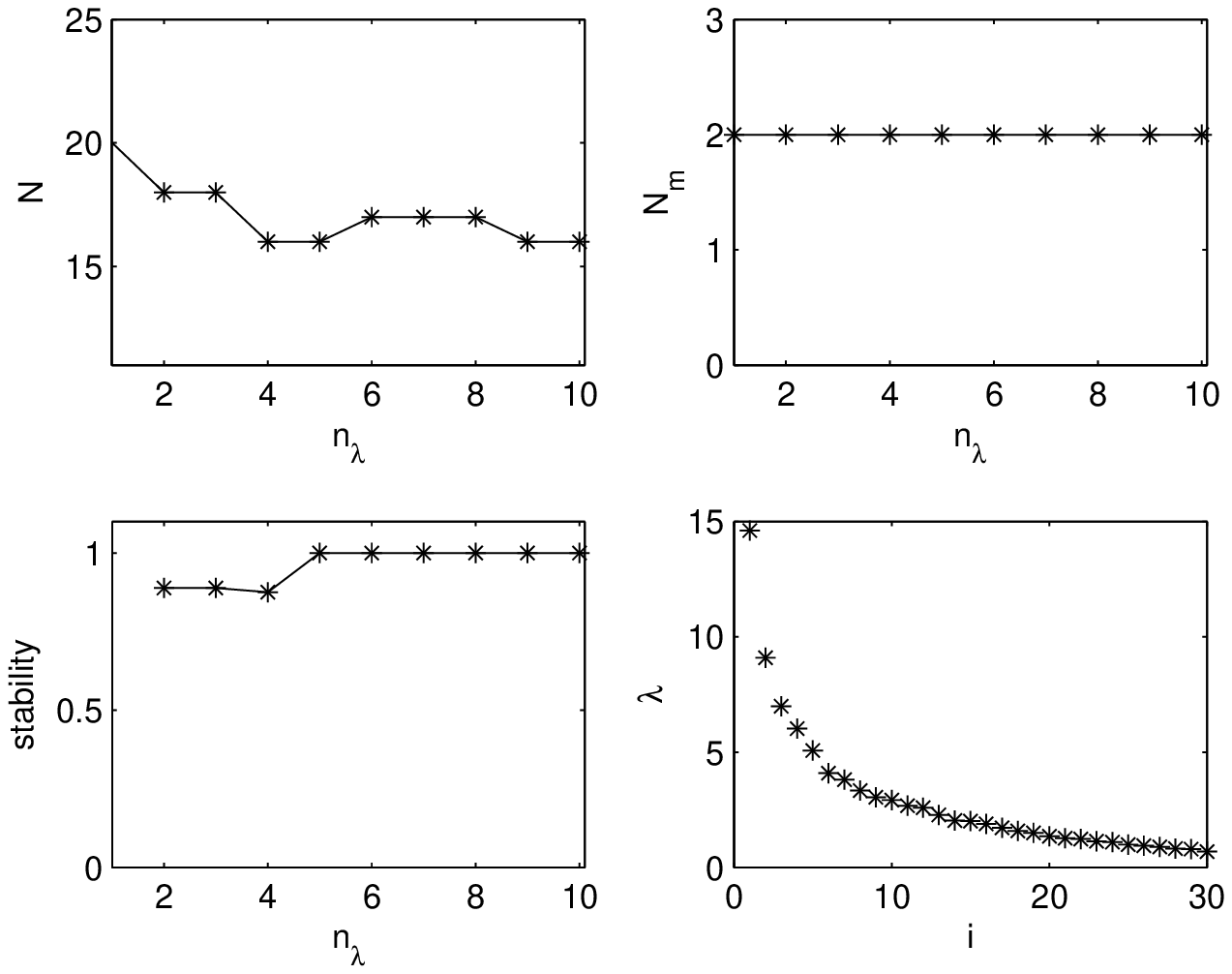,height=7.cm}
\end{center}
\caption{{\small (Top-left) Concerning the genetic application, the
number of selected regions $N$ is plotted versus $n_\lambda$, the
number of modes. (Top-right) The number of modules, obtained by
modularity maximization, of the matrix $c_{ij}$, whose elements
measure the pairwise redundancy. (Bottom-left) The measure of the
stability of the partition, going from $n_\lambda -1$ to
$n_\lambda$, is plotted versus $n_\lambda$. (Bottom-right) The
eigenvalues of the matrix $x^\top x$ are depicted. \label{fig7}}}
\end{figure}
\begin{figure}[ht!]
\begin{center}
\epsfig{file=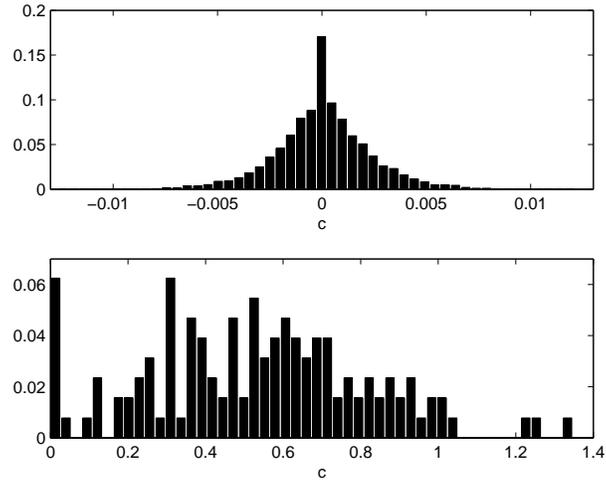,height=7.cm}
\end{center}
\caption{{\small The histogram of the values of the pairwise
redundancy $c_{ij}$, in genetic application (bottom), and choosing
randomly the modes $u$ (top) \label{fig8}}}
\end{figure}

\end{document}